\documentclass{article}
\usepackage[T1]{fontenc}
\usepackage{a4wide}
\usepackage{setspace}
\makeatletter

\providecommand{\LyX}{L\kern-.1667em\lower.25em\hbox{Y}\kern-.125emX\@}

\usepackage[T1]{fontenc}
\makeatletter

\usepackage{amssymb}
\usepackage{amsmath}
\usepackage{amsfonts}

\makeatother
\makeatother
\makeatother

\begin{document}

\title{THE XY MODEL ON THE ONE-DIMENSIONAL SUPERLATTICE: STATIC PROPERTIES}

\author{J. P. de Lima\protect\( ^{a}\protect \) and L. L. Gon\c{c}alves\protect\( ^{b}\protect \)\\
 \protect\( ^{a}\protect \)Departamento de F\'{\i}sica, UFPi, Campus da Ininga,
64049-550\\
 Teresina, Piau\'{\i}, Brazil.\\
 \protect\( ^{b}\protect \)Departamento de F\'{\i}sica, UFCe, Campus do Pici,
C.P. 6030, 60451-970\\
 Fortaleza, Cear\'{a}, Brazil.}

\maketitle
\begin{abstract}
The \( XY \) model (\( s=1/2 \)) on the one-dimensional alternating superlattice
(closed chain) is solved exactly by using a generalized Jordan-Wigner transformation
and the Green function method. Closed expressions are obtained for the excitation
spectrum, the internal energy, the specific heat, the average magnetization
per site, the static susceptibility, \( \chi ^{zz}, \)and the two-spin correlation
function in the field direction at arbitrary temperature. At \( T=0, \) it
is shown that the system presents multiple second order phase transitions induced
by the transverse field, which are associated to the zero energy mode with wave
number equal to \( 0 \) or \( \pi  \). It is also shown that the average magnetization
as a function of the field presents, alternately, regions of plateaux (disordered
phases) and regions of variable magnetization (ordered phases). The static correlation
function presents an oscillating behaviour in the ordered phase and its period
goes to infinity at the critical point. 
\end{abstract}

\section{Introduction}

The experimental development of magnetic superlattices, by using molecular beam
epitaxy technique {[}1-3{]}, has increased the interest in the study of these systems.
Although they are three-dimensional systems, there is a predominance of the
one-dimensional behaviour in their properties, and this is the main reason for
studying one-dimensional superlattices. Therefore, interest has been considerably
increased in the study of spin systems on these lattices.

Among the spin systems the XY-model (\( s \)=1/2), introduced by Lieb et al.
{[}4{]}, occupies a special place, since it can be solved exactly for the homogenous
lattice. Although almost all static and dynamical properties are known for the
model on the homogeneous lattice (see {[}5{]} and references therein), the known
results for non-homogeneous periodic one-dimensional systems are restricted
to the alternating chain {[}6-8{]} and to the excitation spectrum of the general
alternating superlattice {[}9{]}, and its critical behaviour, which has been
obtained by using the position space renormalization group approach {[}10{]}.

In this paper we consider the isotropic XY-model in a transverse field on the
one-dimensional alternating superlattice (closed chain). We solve the model
by introducing a generalized Jordan-Wigner transformation {[}9{]} and by using
the Green function equation of motion tecnhique.

In section 2 we determine the relevant Green functions and present a detailed
discusssion of the excitation spectrum. In section 3 we obtain the internal
energy and the specific heat. The induced magnetization is studied in section
4, and in section 5 we calculate the two-spin correlation function. Finally,
in section 6, we summarize the results and present the main conclusions.

\section{The Excitation Spectrum}

The superlattice that we are going to consider consists of cells composed of
two subcells A and B with \( n_{A} \) and \( n_{B} \) sites, respectively.
The \textit{l-th} unit cell is shown in Fig. 1. The distance \( s \) between
two consecutive sites is taken as unity.

If we assume periodic boundary conditions for a chain with \textit{N} cells,
the Hamiltonian of the \textit{XY} model {[}4{]} can be written in the form
\begin{eqnarray}
H & = & -\frac{1}{2}\sum _{l=1}^{N}\left\{ \sum _{m=1}^{n_{A}-1}J_{A}S_{l,m}^{A^{+}}S_{l,m+1}^{A^{-}}+\sum _{m=1}^{n_{B}-1}J_{B}S_{l,m}^{B^{+}}S_{l,m+1}^{B^{-}}+\right. \nonumber \\
 &  & \nonumber \\
 & + & \left. J\left( S_{l,n_{A}}^{A^{+}}S_{l,1}^{B^{-}}+S_{l,n_{B}}^{B^{+}}S_{l+1,1}^{A^{-}}\right) +h.c.+\right. \nonumber \\
 &  & \nonumber \\
 & + & \left. \sum _{m=1}^{n_{A}}2h_{A}S_{l,m}^{A^{Z}}+\sum _{m=1}^{n_{B}}2h_{B}S_{l,m}^{B^{Z}}\right\} 
\end{eqnarray}
 where \( l \) identifies the cell, \( S^{\pm }=S^{x}\pm iS^{y} \), \( J \)
is the exchange parameter between spins at the interfaces, \( J_{A}(J_{B}) \)
the exchange parameter between spins within the subcell \( A(B), \) and \( h_{A}(h_{B}) \)
is the transverse field within the subcell \( A(B). \) The spin operators can
be expressed in terms of fermion operators using the generalized Jordan-Wigner
transformation {[}9{]} and, by introducing this transformation, the Hamiltonian
can be written in the form 
\begin{eqnarray}
H & = & -\frac{1}{2}\sum ^{N}_{l=1}\left\{ \sum _{m=1}^{n_{A}-1}J_{A}a_{l,m}^{\dag }a_{l,m+1}+\sum _{m=1}^{n_{B}-1}J_{B}b_{l,m}^{\dag }b_{l,m+1}+\right. \nonumber \\
 &  & \nonumber \\
 & + & J\left( a_{l,n_{A}}^{\dag }b_{l,1}+b_{l,n_{B}}^{\dag }a_{l+1,1}\right) +h.c.+\nonumber \\
 &  & \nonumber \\
 & + & \left. \sum _{m=1}^{n_{A}}2h_{A}\left( a_{l,m}^{\dag }a_{l,m}-1\right) +\sum _{m=1}^{n_{B}}2h_{B}\left( b_{l,m}^{\dag }b_{l,m}-1\right) \right\} +\Phi ,
\end{eqnarray}
 where \( a \)'s and \( b \)'s are fermion operators, and \( \Phi  \), given
by 
\begin{equation}
\Phi =\frac{J}{2}\left( b_{N,n_{B}}^{\dag }a_{1,1}+h.c.\right) \exp \left[ i\pi \left( \sum _{l=1}^{N}\sum _{r=1}^{n_{A}}a_{l,r}^{\dag }a_{l,r}+\sum _{l=1}^{N}\sum _{r=1}^{n_{B}}b_{l,r}^{\dag }b_{l,r}\right) \right] ,
\end{equation}

\noindent is a boundary term which will be neglected. As it has been shown {[}11{]},
this boundary term, in the thermodynamic limit, does not affect the excitation
spectrum, the static properties of the system nor the dynamic correlation function
in the field direction. Introducing the Fourier transforms {[}9{]}, 
\begin{eqnarray}
a_{Qk_{1}} & = & \sqrt{\frac{2}{N\left( n_{A}+1\right) }}\sum _{l,m}\exp \left( iQdl\right) \sin \left( mk_{1}\right) a_{l,m},\\
 &  & \nonumber \\
b_{Qk_{2}} & = & \sqrt{\frac{2}{N\left( n_{B}+1\right) }}\sum _{l,m}\exp \left( iQdl\right) \sin \left( mk_{2}\right) b_{l,m},
\end{eqnarray}
 where \( k_{1}=n_{1}\pi /\left( n_{A}+1\right) , \) \( n_{1}=1,2,\ldots ,n_{A}, \)
\( k_{2}=n_{2}\pi /\left( n_{B}+1\right) , \) \( n_{2}=1,2,\ldots ,n_{B}, \)
\( Q=2\pi n/N, \) \( n=1,2,\ldots ,N, \) \( d=n_{A}+n_{B} \) is the size
of the cell, the Hamiltonian can be written as 
\begin{equation}
H=\sum _{Q}H_{Q}+N\left( \frac{n_{A}h_{A}+n_{B}h_{B}}{2}\right) 
\end{equation}
 where \( H_{Q} \) is given by 
\begin{eqnarray}
H_{Q} & = & \sum _{k_{1}}E_{k_{1}}^{A}a_{Qk_{1}}^{\dag }a_{Qk_{1}}+\sum _{k_{2}}E_{k_{2}}^{B}b_{Qk_{2}}^{\dag }b_{Qk_{2}}+\nonumber \\
 &  & \nonumber \\
 & - & \frac{J}{\sqrt{(n_{A}+1)(n_{B}+1)}}\sum _{k_{1},k_{2}}\left\{ \left[ \sin \left( k_{1}n_{A}\right) \sin k_{2}\right. \right. \nonumber \\
 &  & \nonumber \\
 & + & \left. \left. \sin \left( k_{2}n_{B}\right) \sin k_{1}\exp (iQd)a_{Qk_{1}}^{\dag }b_{Qk_{2}}\right] +h.c.\right\} ,
\end{eqnarray}
 with \( E_{k_{1(2)}}^{A(B)}=-J_{A(B)}\cos k_{1(2)}-h_{A(B)} \).

As in the study of the excitation spectrum {[}9{]} we will solve the model by
using the Green function method {[}12{]}. Adopting the notation \( \left\langle \left\langle \Re _{1}(t);\Re _{2}(0)\right\rangle \right\rangle _{r} \)
for the retarded anticommutator function, where \( \Re _{1} \) and \( \Re _{2} \)
are arbitrary operators, and introducing the time Fourier transform defined
as 
\begin{equation}
\left\langle \left\langle \Re _{1}(t);\Re _{2}(0)\right\rangle \right\rangle _{r}=\frac{1}{2\pi }\int _{-\infty }^{\infty }\left\langle \left\langle \Re _{1};\Re _{2}\right\rangle \right\rangle \exp \left( -i\omega t\right) d\omega ,
\end{equation}
 we can write the equation of motion for the Green function \( \ll a_{Qk_{1}};a_{Q^{\prime }k_{1}^{\prime }}^{\dag }\gg  \)
in the form {[}12{]}
\begin{eqnarray}
\left\langle \left\langle a_{Qk_{1}};a_{Q^{\prime }k_{1}^{\prime }}^{\dag }\right\rangle \right\rangle  & = & \delta _{k_{1}k_{1}^{\prime }}\left( \omega -E_{k_{1}}^{A}\right) ^{-1}+\nonumber \\
 &  & \nonumber \\
 & - & \frac{J\left( \omega -E_{k_{1}}^{A}\right) ^{-1}}{\sqrt{(n_{A}+1)(n_{B}+1)}}\sum _{k_{2}}\left[ \sin k_{2}\sin \left( n_{A}k_{1}\right) \right. +\nonumber \\
 &  & \nonumber \\
 & + & \left( iQd\right) \left. \sin k_{1}\sin \left( n_{B}k_{2}\right) \right] \left\langle \left\langle b_{Qk_{2}};a_{Q^{\prime }k_{1}^{\prime }}^{\dag }\right\rangle \right\rangle ,
\end{eqnarray}
 where we have assumed \( \hslash =1. \)

Likewise we can write for \( \ll b_{Qk_{2}};a_{Q^{\prime }k_{1}}^{\dag }\gg  \)
the result 
\begin{eqnarray}
\left\langle \left\langle b_{Qk_{2}};a_{Q^{\prime }k_{1}}^{\dag }\right\rangle \right\rangle  & = & -\frac{J\left( \omega -E_{k_{2}}^{B}\right) ^{-1}}{\sqrt{(n_{A}+1)(n_{B}+1)}}\sum _{k_{1}^{\prime }}\left[ \sin k_{2}\sin \left( n_{A}k_{1}^{\prime }\right) \right. +\nonumber \\
 &  & \nonumber \\
 & + & \left. \exp \left( -iQd\right) \sin k_{1}^{\prime }\sin \left( n_{B}k_{2}\right) \right] \left\langle \left\langle a_{Qk_{1}^{\prime }};a_{Q^{\prime }k_{1}}^{\dag }\right\rangle \right\rangle .
\end{eqnarray}

Eqs. (11) and (12) constitute a closed set which can be easily solved by introducing
the operators \( A_{Q,n} \) and \( B_{Q,n} \) defined by 
\begin{eqnarray}
A_{Q,n} & = & \sqrt{\frac{2}{n_{A}+1}}\sum _{k_{1}}\sin \left( nk_{1}\right) a_{Qk_{1}},\\
 &  & \nonumber \\
B_{Q,n} & = & \sqrt{\frac{2}{n_{B}+1}}\sum _{k_{2}}\sin \left( nk_{2}\right) b_{Qk_{2}}
\end{eqnarray}

Therefore by eliminating the function \( \ll b_{Qk_{2}};a_{Q^{\prime }k_{1}^{\prime }}^{\dag }\gg  \)
from eqs.(11) and (12) and introducing the operator \( A_{Q,n} \) we can find
the result 
\begin{eqnarray}
\left\langle \left\langle A_{Q,m};A_{Q^{\prime },n}^{\dag }\right\rangle \right\rangle  & = & \frac{2}{J}\sqrt{\frac{n_{B}+1}{n_{A}+1}}f_{m,n}^{A}\left( \omega \right) \delta _{Q,Q^{\prime }}+\nonumber \\
 &  & \nonumber \\
 & + & \left[ f_{1,1}^{B}\left( \omega \right) f_{m,n_{A}}^{A}\left( \omega \right) +\exp \left( iQd\right) f_{1,n_{B}}^{B}\left( \omega \right) f_{m,1}^{A}\left( \omega \right) \right] \left\langle \left\langle A_{Q,n_{A}};A_{Q^{\prime },n}^{\dag }\right\rangle \right\rangle +\nonumber \\
 &  & \nonumber \\
 & + & \left[ f_{1,1}^{B}\left( \omega \right) f_{m,1}^{A}\left( \omega \right) +\exp \left( -iQd\right) f_{1,n_{B}}^{B}\left( \omega \right) f_{m,n_{A}}^{A}\left( \omega \right) \right] \left\langle \left\langle A_{Q,1};A_{Q^{\prime },n}^{\dag }\right\rangle \right\rangle \nonumber \\
 &  & 
\end{eqnarray}
 where 
\begin{eqnarray}
f_{m,n}^{A}\left( \omega \right)  & \equiv  & \frac{J}{\sqrt{\left( n_{A}+1\right) \left( n_{B}+1\right) }}\sum _{k_{1}}\frac{\sin \left( mk_{1}\right) \sin \left( nk_{1}\right) }{\omega -E_{k_{1}}^{A}},\\
 &  & \nonumber \\
f_{m,n}^{B}\left( \omega \right)  & \equiv  & \frac{J}{\sqrt{\left( n_{A}+1\right) \left( n_{B}+1\right) }}\sum _{k_{2}}\frac{\sin \left( mk_{2}\right) \sin \left( nk_{2}\right) }{\omega -E_{k_{2}}^{B}}.
\end{eqnarray}

From eq. (15) we can find immediately that \( \ll A_{Q,m};A_{Q^{\prime },n}^{\dag }\gg  \)
is given by 
\begin{eqnarray}
\left\langle \left\langle A_{Q,m};A_{Q^{\prime },n}^{\dag }\right\rangle \right\rangle  & = & \frac{2}{J}\sqrt{\frac{n_{B}+1}{n_{A}+1}}\delta _{Q,Q^{\prime }}\left\{ f_{m,n}^{A}\left( \omega \right) \right. +\frac{1}{R_{Q}\left( \omega \right) }\times \nonumber \\
 &  & \nonumber \\
 & \times  & \left\{ \left[ f_{1,1}^{B}\left( \omega \right) f_{m,n_{A}}^{A}\left( \omega \right) +\exp \left( iQd\right) f_{1,n_{B}}^{B}\left( \omega \right) f_{m,1}^{A}\left( \omega \right) \right] \right. \times \nonumber \\
 &  & \nonumber \\
 & \times  & \left[ f_{1,n}^{A}\left( \omega \right) \left( f_{1,n_{A}}^{A}\left( \omega \right) f_{1,1}^{B}\left( \omega \right) +f_{1,1}^{A}\left( \omega \right) f_{1,n_{B}}^{B}\left( \omega \right) \exp \left( -iQd\right) \right) \right. +\nonumber \\
 &  & \nonumber \\
 & + & \left. f_{n,n_{A}}^{A}\left( \omega \right) \left( 1-f_{1,1}^{A}\left( \omega \right) f_{1,1}^{B}\left( \omega \right) -f_{1,n_{A}}^{A}\left( \omega \right) f_{1,n_{B}}^{B}\left( \omega \right) \exp \left( -iQd\right) \right) \right] +\nonumber \\
 &  & \nonumber \\
 & + & \left[ f_{1,1}^{B}\left( \omega \right) f_{m,1}^{A}\left( \omega \right) +\exp \left( -iQd\right) f_{1,n_{B}}^{B}\left( \omega \right) f_{m,n_{A}}^{A}\left( \omega \right) \right] \times \nonumber \\
 &  & \nonumber \\
 & \times  & \left[ f_{n,n_{A}}^{A}\left( \omega \right) \left( f_{1,n_{A}}^{A}\left( \omega \right) f_{1,1}^{B}\left( \omega \right) +f_{1,1}^{A}\left( \omega \right) f_{1,n_{B}}^{B}\left( \omega \right) \exp \left( iQd\right) \right) \right. +\nonumber \\
 &  & \nonumber \\
 & + & \left. \left. \left. f_{1,n}^{A}\left( \omega \right) \left( 1-f_{1,1}^{A}\left( \omega \right) f_{1,1}^{B}\left( \omega \right) -f_{1,n_{A}}^{A}\left( \omega \right) f_{1,n_{B}}^{B}\left( \omega \right) \exp \left( iQd\right) \right) \right] \right\} \right\} .\nonumber \\
 &  & 
\end{eqnarray}
 The details of the calculation can be found in ref. {[}13{]}.

By a similar procedure we can write the set of equations 
\begin{eqnarray}
\left\langle \left\langle b_{Qk_{2}};b_{Q^{\prime }k_{2}^{\prime }}^{\dag }\right\rangle \right\rangle  & = & \delta _{k_{2}k_{2}^{\prime }}\left( \omega -E_{k_{2}}^{B}\right) ^{-1}+\nonumber \\
 &  & \nonumber \\
 & - & \frac{J\left( \omega -E_{k_{2}}^{B}\right) ^{-1}}{\sqrt{(n_{A}+1)(n_{B}+1)}}\sum _{k_{1}}\left[ \sin k_{2}\sin \left( n_{A}k_{1}\right) \right. +\nonumber \\
 &  & \nonumber \\
 & + & \left. \exp \left( -iQd\right) \sin k_{1}\sin \left( n_{B}k_{2}\right) \right] \left\langle \left\langle a_{Qk_{1}};b_{Q^{\prime }k_{2}^{\prime }}^{\dag }\right\rangle \right\rangle ,
\end{eqnarray}
 
\begin{eqnarray}
 &  & \nonumber \\
\left\langle \left\langle a_{Qk_{1}};b_{Q^{\prime }k_{2}^{\prime }}^{\dag }\right\rangle \right\rangle  & = & -\frac{J\left( \omega -E_{k_{1}}^{A}\right) ^{-1}}{\sqrt{(n_{A}+1)(n_{B}+1)}}\sum _{k_{2}}\left[ \sin k_{2}\sin \left( n_{A}k_{1}\right) \right. +\nonumber \\
 &  & \nonumber \\
 & + & \left. \exp \left( iQd\right) \sin k_{1}\sin \left( n_{B}k_{2}\right) \right] \left\langle \left\langle b_{Qk_{2}};b_{Q^{\prime }k_{2}^{\prime }}^{\dag }\right\rangle \right\rangle ,
\end{eqnarray}
 which can be solved by eliminating the function \( \ll a_{Qk_{1}};b_{Q^{\prime }k_{2}^{\prime }}^{\dag }\gg  \)
in the previous equation and by introducing the operator \( B_{Q,n}. \)This
procedure leads to the set of equations 
\begin{eqnarray}
\left\langle \left\langle B_{Q,m};B_{Q^{\prime },n}^{\dag }\right\rangle \right\rangle  & = & \frac{2}{J}\sqrt{\frac{n_{A}+1}{n_{B}+1}}f_{m,n}^{B}\left( \omega \right) \delta _{Q,Q^{\prime }}+\nonumber \\
 &  & \nonumber \\
 & + & \left[ f_{1,1}^{A}\left( \omega \right) f_{m,n_{B}}^{B}\left( \omega \right) +\exp \left( iQd\right) f_{1,n_{A}}^{A}\left( \omega \right) f_{m,1}^{B}\left( \omega \right) \right] \left\langle \left\langle B_{Q,n_{B}};B_{Q^{\prime },n}^{\dag }\right\rangle \right\rangle +\nonumber \\
 &  & \nonumber \\
 & + & \left[ f_{1,1}^{A}\left( \omega \right) f_{m,1}^{B}\left( \omega \right) +\exp \left( -iQd\right) f_{1,n_{A}}^{A}\left( \omega \right) f_{m,n_{B}}^{B}\left( \omega \right) \right] \left\langle \left\langle B_{Q,1};B_{Q^{\prime },n}^{\dag }\right\rangle \right\rangle ,\nonumber \\
 &  & 
\end{eqnarray}
 and, as we can see, it can be obtained from eq. (15 ), provided we make the
substitution A\( \rightarrow  \)B and B\( \rightarrow  \)A. Therefore \( \ll B_{Q,m};B_{Q^{\prime },n}^{\dag }\gg  \)
is obtained from eq. (18) by introducing the previous transformation.

The excitation spectrum is given by the poles of \( \ll A_{Q,m};A_{Q^{\prime },n}^{\dag }\gg  \),
or \( \ll B_{Q,m};B_{Q^{\prime },n}^{\dag }\gg  \), and corresponds to the
solution of the equation \( R_{Q}\left( \omega \right) =0 \), which is explicitly
given by {[}9{]}
\begin{eqnarray}
1 & - & 2f_{1,n_{A}}^{A}\left( \omega \right) f_{1,n_{B}}^{B}\left( \omega \right) \cos \left( Qd\right) -2f_{1,1}^{A}\left( \omega \right) f_{1,1}^{B}\left( \omega \right) +\nonumber \\
 &  & \nonumber \\
 & + & \left[ \left( f_{1,1}^{A}\left( \omega \right) \right) ^{2}-\left( f_{1,n_{A}}^{A}\left( \omega \right) \right) ^{2}\right] \left[ \left( f_{1,1}^{B}\left( \omega \right) \right) ^{2}-\left( f_{1,n_{B}}^{B}\left( \omega \right) \right) ^{2}\right] =0.
\end{eqnarray}

As mentioned in our previous paper {[}9{]}, it should be noted that the values
\( \omega =E_{k_{1}}^{A} \) and \( \omega =E_{k_{2}}^{B} \) are not poles
of \( \ll A_{Q,m};A_{Q^{\prime },n}^{\dag }\gg  \), since in this limit the
function is finite. Therefore, as expected, the spectrum of each subcell does
not coincide with spectrum of the superlattice, and contains \( n_{A}+n_{B} \)
branches {[}9{]}.

The general solution of this equation is determined numerically, although analytical
solutions can be found for some special cases. For instance for \( n_{A}=n_{B}=2,h_{A}=h_{B}=h, \)
we can express explicitly the solution in the form 
\begin{equation}
\omega _{Q}=-h\pm \frac{1}{\sqrt{2}}\sqrt{c\pm \sqrt{g\left( Qd\right) }},
\end{equation}
 where 
\begin{eqnarray}
c & \equiv  & \frac{J^{2}}{2}+\frac{1}{4}\left( J_{A}^{2}+J_{B}^{2}\right) ,\\
 &  & \nonumber \\
g\left( Qd\right)  & \equiv  & \frac{1}{2}J^{2}J_{A}J_{B}\cos \left( Qd\right) +\frac{J^{2}}{4}\left( J_{A}^{2}+J_{B}^{2}\right) +\frac{1}{16}\left( J_{A}^{2}-J_{B}^{2}\right) ^{2},
\end{eqnarray}

\noindent which for \( J_{A}=J_{B} \) reproduces the known results {[}6,7{]}.

In the homogeneous medium limit, the spectrum obtained from the eq. (22) presents
\( n_{A}+n_{B} \) branches, which correspond to a (\( n_{A}+n_{B}-1 \))-folding
of the spectrum. As an example of this limit we present in Fig. 2 the excitation
spectrum for \( n_{A}=2, \) \( n_{B}=3, \) \( J_{A}=J_{B}=J=1 \) and \( h_{A}=h_{B} \)
\( =2 \).

For zero field and \( n_{A}+n_{B} \) odd, there is a zero energy mode with
wave number different from zero, as shown in Fig. 3 for \( n_{A}=2, \) \( n_{B}=3, \)
\( J_{A}=2,J_{B}=3 \) and \( J=1 \). On the other hand, for \( n_{A}+n_{B} \)
even this mode is not present, and for each wave number \( Q \) there are symmetrical
solutions \( +\omega _{Q} \) and \( -\omega _{Q} \), as can be seen in Fig.
4 for \( n_{A}=n_{B}=4, \) \( J_{A}=0.3,J_{B}=3 \) and \( J=1 \).

As it has already been shown {[}14{]}, the effect of a homogeneous field, \( h_{A}=h_{B}=h, \)is
to shift the zero field spectrum. This can also be shown directly from eq. (22)
and, consequently, the existence of a mode of zero energy will depend on the
strength of the field.

As we can also see in Fig. 4, the extreme bands of the spectrum are very narrow,
and this is related to the difference between the exchange parameters of subcells
\( A \) and \( B \). As the difference between the parameters of the media
\( A \) and \( B \) decreases, the dispersion increases, and the gaps tend
to zero. This is the expected behaviour, since we have a maximum dispersion
in the homogeneous limit, as shown in Fig. 2. In all cases presented the number
of branches is equal to the number of sites per cell.

Again, we note that the spectrum can also be calculated exactly by using the
position space renormalization group approach {[}10{]}, and approximately by
using a transfer matrix method {[}15{]}. Although the latter is an approximate
method, we have shown that it reproduces numerically the exact result {[}5,8{]}.

\section{The Internal Energy and the Specific Heat}

The operator \( H_{Q} \) defined in eq. (9) can be written in terms of operators
\( A_{Q,n} \) and \( B_{Q,n} \) in the form 
\begin{eqnarray}
H_{Q} & = & \frac{2}{n_{A}+1}\sum _{k_{1}}E_{k_{1}}^{A}\left[ \sum _{n=1}^{n_{A}}\sin \left( nk_{1}\right) A_{Q,n}^{\dag }\right] \left[ \sum _{n=1}^{n_{A}}\sin \left( nk_{1}\right) A_{Q,n}\right] +\nonumber \\
 &  & \nonumber \\
 & + & \frac{2}{n_{B}+1}\sum _{k_{2}}E_{k_{2}}^{B}\left[ \sum _{n=1}^{n_{B}}\sin \left( nk_{2}\right) B_{Q,n}^{\dag }\right] \left[ \sum _{n=1}^{n_{B}}\sin \left( nk_{2}\right) B_{Q,n}\right] +\nonumber \\
 &  & \nonumber \\
 & - & \frac{J}{2}\left[ A_{Q,n_{A}}^{\dag }B_{Q,1}+\exp \left( iQd\right) A_{Q,1}^{\dag }B_{Q,n_{B}}+h.c.\right] .
\end{eqnarray}
 From this equation we can see that the internal energy, \( \left\langle H\right\rangle =\sum _{Q}\left\langle H_{Q}\right\rangle  \),
can be calculated from the Green function \( G_{Q}\left( \omega \right)  \)
given by 
\begin{equation}
G_{Q}\left( \omega \right) =G1_{Q}\left( \omega \right) +G2_{Q}\left( \omega \right) +G3_{Q}\left( \omega \right) ,
\end{equation}
 where

\begin{eqnarray}
G1_{Q}\left( \omega \right)  & = & \frac{2}{n_{A}+1}\sum _{k_{1}}E_{k_{1}}^{A}\left[ \sum _{m=1}^{n_{A}}\sum _{n=1}^{n_{A}}\sin \left( mk_{1}\right) \sin \left( nk_{1}\right) \left\langle \left\langle A_{Q,m};A_{Q,n}^{\dag }\right\rangle \right\rangle \right] \nonumber \\
 &  & \\
G2_{Q}\left( \omega \right)  & = & \frac{2}{n_{B}+1}\sum _{k_{2}}E_{k_{2}}^{B}\left[ \sum _{m=1}^{n_{B}}\sum _{n=1}^{n_{B}}\sin \left( mk_{2}\right) \sin \left( nk_{2}\right) \left\langle \left\langle B_{Q,m};B_{Q,n}^{\dag }\right\rangle \right\rangle \right] \nonumber \\
 &  & \\
G3_{Q}\left( \omega \right)  & = & -\frac{J}{2}\left[ \left\langle \left\langle B_{Q1};A_{Q,n_{A}}^{\dag }\right\rangle \right\rangle +\exp \left( iQd\right) \left\langle \left\langle B_{Qn_{B}};A_{Q,1}^{\dag }\right\rangle \right\rangle +\right. \nonumber \\
 &  & \nonumber \\
 & + & \left. \left\langle \left\langle A_{Qn_{A}};B_{Q,1}^{\dag }\right\rangle \right\rangle +\exp \left( -iQd\right) \left\langle \left\langle A_{Q1};B_{Q,n_{B}}^{\dag }\right\rangle \right\rangle \right] .
\end{eqnarray}

The Green functions \( \ll B_{Q1};A_{Q,n_{A}}^{\dag }\gg  \), \( \ll B_{Qn_{B}};A_{Q,1}^{\dag }\gg  \),
\( \ll A_{Qn_{A}};B_{Q,1}^{\dag }\gg  \) and \( \ll A_{Q1};B_{Q,n_{B}}^{\dag }\gg  \)
can, with the aid of eqs. (11-14) and (19,20), be written in terms of the functions
\( \ll A_{Q,m};A_{Q^{\prime },n}^{\dag }\gg  \) and \( \ll B_{Q,m};B_{Q^{\prime },n}^{\dag }\gg  \),
and are given by

\begin{eqnarray}
\left\langle \left\langle A_{Q,1};B_{Q^{\prime },n_{B}}^{\dag }\right\rangle \right\rangle  & = & -\sqrt{\frac{n_{B}+1}{n_{A}+1}}\left[ f_{1,n_{A}}^{A}\left( \omega \right) \left\langle \left\langle B_{Q,1};B_{Q^{\prime },n_{B}}^{\dag }\right\rangle \right\rangle \right. +\nonumber \\
 &  & \nonumber \\
 & + & \left. \exp \left( iQd\right) f_{1,1}^{A}\left( \omega \right) \left\langle \left\langle B_{Q,n_{B}};B_{Q^{\prime },n_{B}}^{\dag }\right\rangle \right\rangle \right] ,
\end{eqnarray}
 
\begin{eqnarray}
\left\langle \left\langle A_{Q,n_{A}};B_{Q^{\prime },1}^{\dag }\right\rangle \right\rangle  & = & -\sqrt{\frac{n_{B}+1}{n_{A}+1}}\left[ f_{1,1}^{A}\left( \omega \right) \left\langle \left\langle B_{Q,1};B_{Q^{\prime },1}^{\dag }\right\rangle \right\rangle \right. +\nonumber \\
 &  & \nonumber \\
 & + & \left. \exp \left( iQd\right) f_{1,n_{A}}^{A}\left( \omega \right) \left\langle \left\langle B_{Q,n_{B}};B_{Q^{\prime },1}^{\dag }\right\rangle \right\rangle \right] ,
\end{eqnarray}
 
\begin{eqnarray}
\left\langle \left\langle B_{Q,n_{B}};A_{Q^{\prime },1}^{\dag }\right\rangle \right\rangle  & = & -\sqrt{\frac{n_{A}+1}{nB+1}}\left[ f_{1,n_{B}}^{B}\left( \omega \right) \left\langle \left\langle A_{Q,n_{A}};A_{Q^{\prime },1}^{\dag }\right\rangle \right\rangle \right. +\nonumber \\
 &  & \nonumber \\
 & + & \left. \exp \left( -iQd\right) f_{1,1}^{B}\left( \omega \right) \left\langle \left\langle A_{Q,1};A_{Q^{\prime },1}^{\dag }\right\rangle \right\rangle \right] ,
\end{eqnarray}

\noindent 
\begin{eqnarray}
\left\langle \left\langle B_{Q1};A_{Q^{\prime },n_{A}}^{\dag }\right\rangle \right\rangle  & = & -\sqrt{\frac{n_{A}+1}{nB+1}}\left[ f_{1,1}^{B}\left( \omega \right) \left\langle \left\langle A_{Q,n_{A}};A_{Q^{\prime },n_{A}}^{\dag }\right\rangle \right\rangle \right. +\nonumber \\
 &  & \nonumber \\
 & + & \left. \exp \left( -iQd\right) f_{1,n_{B}}^{B}\left( \omega \right) \left\langle \left\langle A_{Q,1};A_{Q^{\prime },n_{A}}^{\dag }\right\rangle \right\rangle \right] .
\end{eqnarray}

Then by using eqs. (18), and (31-34) in eqs. (28-29) we obtain the Green functions,
\( G1_{Q}\left( \omega \right) , \) \( G2_{Q}\left( \omega \right)  \) and
\( G3_{Q}\left( \omega \right) , \) which are given by {[}13{]}
\begin{eqnarray}
G1_{Q}\left( \omega \right)  & = & \sum _{k_{1}}\frac{E_{k_{1}}^{A}}{\left( \omega -E_{k_{1}}^{A}\right) }+\frac{J\overline{c}}{R_{Q}\left( \omega \right) }\sum _{k_{1}}\frac{E_{k_{1}}^{A}}{\left( \omega -E_{k_{1}}^{A}\right) ^{2}}\nonumber \\
 &  & \nonumber \\
 & \times  & \left\{ \sin ^{2}k_{1}\left[ f_{1,1}^{B}\left( \omega \right) -f_{1,1}^{A}\left( \omega \right) \left( \left( f_{1,1}^{B}\left( \omega \right) \right) ^{2}-\left( f_{1,n_{B}}^{B}\left( \omega \right) \right) ^{2}\right) \right] +\right. \nonumber \\
 &  & \nonumber \\
 & + & \sin k_{1}\sin \left( n_{A}k_{1}\right) \left[ f_{1,n_{B}}^{B}\left( \omega \right) \cos \left( Qd\right) +\right. \nonumber \\
 &  & \nonumber \\
 & + & \left. \left. f_{1,n_{A}}^{A}\left( \omega \right) \left( \left( f_{1,1}^{B}\left( \omega \right) \right) ^{2}-\left( f_{1,n_{B}}^{B}\left( \omega \right) \right) ^{2}\right) \right] \right\} ,
\end{eqnarray}
 
\begin{eqnarray}
G3_{Q}\left( \omega \right)  & = & \frac{2}{R_{Q}\left( \omega \right) }\left\{ 1-R_{Q}\left( \omega \right) -\left[ \left( f_{1,1}^{A}\left( \omega \right) \right) ^{2}-\left( f_{1,n_{A}}^{A}\left( \omega \right) \right) ^{2}\right] \right. \nonumber \\
 &  & \nonumber \\
 & \times  & \left. \left[ \left( f_{1,1}^{B}\left( \omega \right) \right) ^{2}-\left( f_{1,n_{B}}^{B}\left( \omega \right) \right) ^{2}\right] \right\} ,
\end{eqnarray}
 and \( G2_{Q}\left( \omega \right)  \) can be obtained from \( G1_{Q}\left( \omega \right)  \)
by introducing the transformations \( k_{1}\rightarrow  \) \( k_{2}, \) \( A \)
\( \rightarrow B \) and \( B\rightarrow  \) \( A. \) From these results we
can show that the expression for \( G_{Q}\left( \omega \right)  \) has the
form 
\begin{equation}
G_{Q}\left( \omega \right) =\frac{1}{R_{Q}\left( \omega \right) }\left[ \mathcal{G}\left( f_{1,1}^{A}\left( \omega \right) ,f_{1,n_{A}}^{A}\left( \omega \right) ,f_{1,1}^{B}\left( \omega \right) ,f_{1,n_{B}}^{B}\left( \omega \right) \right) \right] .
\end{equation}

Then, the internal energy, \( \left\langle H\right\rangle =\sum _{Q}\left\langle H_{Q}\right\rangle , \)
can be obtained from the expression {[}7{]}

\begin{equation}
\left\langle H\right\rangle =\frac{1}{\pi }\sum _{Q}\int\limits _{-\infty }^{\infty }\frac{Im\left( G_{Q}\left( \omega \right) \right) }{e^{\beta \omega }+1}d\omega ,
\end{equation}

\noindent where, as usual, \( \beta =1/k_{B}T. \)

By using the Dirac identity 
\begin{equation}
\frac{1}{x-x_{0}\pm i\varepsilon }\equiv \mathcal{P}\left( \frac{1}{x-x_{0}}\right) \mp i\pi \delta \left( x-x_{0}\right) ,
\end{equation}
 we can show immediately that \( Im\left( G_{Q}\left( \omega \right) \right)  \)
can be written in the form 
\begin{equation}
Im\left( G_{Q}\left( \omega \right) \right) =\sum _{r}F_{Q,r}\delta \left( \omega -\omega _{Q,r}\right) ,
\end{equation}
 where 
\begin{equation}
F_{Q,r}=\frac{R_{Q}\left( \omega _{Q,r}\right) G_{Q}\left( \omega _{Q,r}\right) }{R_{Q}^{\prime }\left( \omega _{Q,r}\right) },
\end{equation}
 where \( r \) labels the branches of the spectrum and \( R_{Q}^{\prime }\left( \omega \right)  \)
is equal to \( dR_{Q}\left( \omega \right) /d\omega . \)

The specific heat is obtained from eq. (38) and can be written in the form 
\begin{equation}
C=\frac{1}{N}\frac{d\left\langle H\right\rangle }{dT}=\frac{k_{B}\beta ^{2}}{N\pi }\sum _{Q}\int\limits _{-\infty }^{\infty }\frac{\omega e^{\beta \omega }Im\left( G_{Q}\left( \omega \right) \right) }{\left( e^{\beta \omega }+1\right) ^{2}}d\omega .
\end{equation}

The behaviour of the internal energy as a function of temperature does not present
any remarkable difference as we change the superlattice parameters. A typical
result is shown in Fig. 5 where the internal energy is shown as a function of
temperature \( T^{*}(T^{*}\equiv k_{B}T) \) for \( n_{A}=n_{B}=10, \) \( J_{A}=2,J_{B}=3, \)
\( J=1 \), \( h_{A}=h_{B} \) \( =1.5. \)

On the other hand, the behaviour of the specific heat as a function of temperature
is very susceptible to these parameters. It can present a single peak or a double
peak, as it can be seen in Figs. 6 and 7. This important feature, namely, the
appearance of the double peak, is a consequence of the packing of the branches
of the excitation spectrum, which is strongly dependent on the interaction parameters.
As we increase the field, we move the spectrum downwards and this has the effect
of suppressing one peak of the specific heat. This is shown in Fig. 8, where
we have used the same lattice parameters of Fig. 7, and a larger field.

In Figs. 9 and 10 we present the low temperature behaviour of the specific heat
shown in Figs. 7 and 8, respectively. For the results shown in Fig. 9, the excitation
spectrum has a zero energy mode contrary to the case shown in Fig. 10, where
it is not present, as can be verified by using the analytical solution presented
in eq. (23). As expected, the derivative of the specific heat, \( dC/dT^{*}, \)
at \( T=0, \) is equal to zero only when there is a zero mode energy in the
spectrum.

\section{The Induced Magnetization and the Susceptibility \protect\( \chi ^{zz}\protect \)}

The local induced magnetization which corresponds to the average value of \( S_{j,m}^{A^{z}} \)
and \( S_{j,m}^{B^{z}}, \) is given by 
\begin{equation}
\left\langle S_{j,m}^{A^{z}}\right\rangle =\left\langle a_{j,m}^{\dag }a_{j,m}\right\rangle -\frac{1}{2}\quad \text {and\quad }\left\langle S_{j,m}^{B^{z}}\right\rangle =\left\langle b_{j,m}^{\dag }b_{j,m}\right\rangle -\frac{1}{2},
\end{equation}
 which can be determined from the Green functions \( \ll a_{j,m};a_{j,m}^{\dag }\gg  \)
and \( \ll b_{j,m};b_{j,m}^{\dag }\gg  \), respectively. In terms of the operators
\( A_{Q,m} \) and \( B_{Q,m} \) these functions can be written in the form
\begin{equation}
\left\langle \left\langle a_{j,m};a_{j,m}^{\dag }\right\rangle \right\rangle =\frac{1}{N}\sum _{Q,Q^{\prime }}\exp \left[ -i\left( Q-Q^{\prime }\right) d\right] \left\langle \left\langle A_{Q,m};A_{Q^{\prime },m}^{\dag }\right\rangle \right\rangle 
\end{equation}
 and 
\begin{equation}
\left\langle \left\langle b_{j,m};b_{j,m}^{\dag }\right\rangle \right\rangle =\frac{1}{N}\sum _{Q,Q^{\prime }}\exp \left[ -i\left( Q-Q^{\prime }\right) d\right] \left\langle \left\langle B_{Q,m};B_{Q^{\prime },m}^{\dag }\right\rangle \right\rangle .
\end{equation}
 Then introducing in eq. (44) the function \( \ll A_{Q,m};A_{Q^{\prime },m}^{\dag }\gg  \)
, which is given in eq. (18), we can write \( \ll a_{j,m};a_{j,m}^{\dag }\gg  \)
in the form 
\begin{eqnarray}
\left\langle \left\langle a_{j,m};a_{j,m}^{\dag }\right\rangle \right\rangle  & = & \frac{2}{NJ}\sqrt{\frac{n_{B}+1}{n_{A}+1}}\sum _{Q}\left\{ f_{m,m}^{A}\left( \omega \right) \right. +\nonumber \\
 &  & \nonumber \\
 & + & \frac{1}{R_{Q}\left( \omega \right) }\left[ f_{1,m}^{A}\left( \omega \right) f_{m,n_{A}}^{A}\left( \omega \right) \right. \left( 2f_{1,n_{B}}^{B}\left( \omega \right) \cos (Qd)+2f_{1,n_{A}}^{A}\left( \omega \right) \right. \nonumber \\
 &  & \nonumber \\
 & \times  & \left. \left( \left( f_{1,1}^{B}\left( \omega \right) \right) ^{2}-\left( f_{1,n_{B}}^{B}\left( \omega \right) \right) ^{2}\right) \right) +\left( \left( f_{1,m}^{A}\left( \omega \right) \right) ^{2}-\left( f_{m,n_{A}}^{A}\left( \omega \right) \right) ^{2}\right) \nonumber \\
 &  & \nonumber \\
 & \times  & \left. \left. \left( f_{1,1}^{B}\left( \omega \right) +f_{1,1}^{A}\left( \omega \right) \right) \left( \left( f_{1,1}^{B}\left( \omega \right) \right) ^{2}-\left( f_{1,n_{B}}^{B}\left( \omega \right) \right) ^{2}\right) \right] \right\} .
\end{eqnarray}
 The function \( \ll b_{j,m};b_{j,m}^{\dag }\gg  \) is obtained from \( \ll a_{j,m};a_{j,m}^{\dag }\gg  \)
by introducing the transformations \( f^{A}\rightarrow f^{B} \), \( f^{B}\rightarrow f^{A} \),
\( n_{A}\rightarrow n_{B} \) and \( n_{B}\rightarrow n_{A}. \) From this result
we can write 
\begin{equation}
\left\langle a_{j,m}^{\dag }a_{j,m}\right\rangle =\frac{1}{\pi }\int _{-\infty }^{\infty }\frac{Im\left( \left\langle \left\langle a_{j,m};a_{j,m}^{\dag }\right\rangle \right\rangle \right) }{e^{\beta \omega }+1}d\omega ,
\end{equation}
 which, in the thermodynamic limit, allows us to obtain \( \left\langle S_{j,m}^{A^{z}}\right\rangle  \)
from the equation 
\begin{equation}
\left\langle S_{j,m}^{A^{z}}\right\rangle =-\frac{1}{2}+\frac{1}{2\pi }\sum _{r}\int _{0}^{2\pi }\frac{F\left( q,\omega _{q,r},h\right) }{\left( e^{\beta \omega _{q,r}}+1\right) \frac{dR_{Q}\left( \omega \right) }{d\omega }\left| _{\omega _{q,r}}\right. }dq,
\end{equation}
 where 
\begin{eqnarray}
F\left( q,\omega _{q,r},h\right)  & = & \frac{2}{J}\sqrt{\frac{n_{B}+1}{n_{A}+1}}\left\{ f_{1,m}^{A}\left( \omega \right) f_{m,n_{A}}^{A}\left( \omega \right) \left[ 2f_{1,n_{B}}^{B}\left( \omega \right) \cos (q)+2f_{1,n_{A}}^{A}\left( \omega \right) \right. \times \right. \nonumber \\
 &  & \nonumber \\
 & \times  & \left. \left( \left( f_{1,1}^{B}\left( \omega \right) \right) ^{2}-\left( f_{1,n_{B}}^{B}\left( \omega \right) \right) ^{2}\right) \right] +\left( \left( f_{1,m}^{A}\left( \omega \right) \right) ^{2}-\left( f_{m,n_{A}}^{A}\left( \omega \right) \right) ^{2}\right) \times \nonumber \\
 &  & \nonumber \\
 & \times  & \left. \left. \left( f_{1,1}^{B}\left( \omega \right) +f_{1,1}^{A}\left( \omega \right) \right) \left( \left( f_{1,1}^{B}\left( \omega \right) \right) ^{2}-\left( f_{1,n_{B}}^{B}\left( \omega \right) \right) ^{2}\right) \right] \right\} ,
\end{eqnarray}
 and where \( q\equiv Qd \) . The local induced magnetization in the subcell
B, \( \left\langle S_{j,m}^{B^{z}}\right\rangle , \)is determined by following
the same procedure and by using the function \( \ll b_{j,m};b_{j,m}^{\dag }\gg  \).

The average induced magnetization in the cell, \( \left\langle S_{cel}^{z}\right\rangle  \),
is defined as 
\begin{equation}
\left\langle S_{cel}^{z}\right\rangle =\frac{1}{N\left( n_{A}+n_{B}\right) }\left[ \sum _{l=1}^{N}\left( \sum _{m=1}^{n_{A}}\left\langle S_{l,m}^{A^{z}}\right\rangle +\sum _{m=1}^{n_{B}}\left\langle S_{l,m}^{B^{z}}\right\rangle \right) \right] 
\end{equation}

\noindent and from this expression, for the case \( h_{A}=h_{B}\equiv h \),
we obtain susceptibility \( \chi ^{zz} \), which is given by 
\begin{equation}
\chi ^{zz}=\frac{d}{dh}\left\langle S_{cel}^{z}\right\rangle .
\end{equation}

Figure 11 shows the average magnetization in the cell and the susceptibility as
a function of\textit{\ {}}\( h \) at \( T=0 \), for \( J=1 \), \( J_{A}=2 \),
\( J_{B}=3, \) \( h_{A}=h_{B}=h \) and \( n_{A}=n_{B}=2. \) As can be seen,
the magnetization presents three plateaus which are limited by four critical
fields, \( h_{c} \), which correspond to the singular points of the susceptibility
\( \chi ^{zz} \) (\( \chi ^{zz}\rightarrow \infty  \) when \( h\rightarrow h_{c}). \)
These four singularities correspond to the modes of zero energy with wave number
equal to \( 0 \) or \( \pi , \) which limit each energy band. For this case,
the critical fields can be obtained exactly from eq. (23) and are given by
\begin{eqnarray}
h_{1c} & = & \frac{5-\sqrt{5}}{4}\cong 0.691,\nonumber \\
h_{2c} & = & \frac{\sqrt{29}-1}{4}\cong 1.096,\nonumber \\
h_{3c} & = & \frac{\sqrt{29}+1}{4}\cong 1.596,\nonumber \\
h_{4c} & = & \frac{\sqrt{5}+5}{4}\cong 1.809.
\end{eqnarray}

It should be noted that the plateaus, where the susceptibility goes to zero,
correspond to the gaps in the spectrum.

Figure 12 shows the average induced magnetization , at \( T=0 \), for \( J=1 \),
\( J_{A}=2 \), \( J_{B}=3, \) \( h_{A}=h_{B}=h \) and \( n_{A}=2,n_{B}=3, \)
as a function of h. In this case, since the number of sites per cell is odd,
there is a zero energy mode even for \( h=0, \) and consequently there is no
zero magnetization plateaux as in the case shown in Fig. 11. The critical behaviour
is, naturally, also present when we have an inhomogeneous field, \( h_{A}=1.25h_{B}=h, \)
and this is shown in Fig. 13 for a larger unit cell, \( n_{A}=2,n_{B}=3. \)

In Fig. 14 we present the magnetization for finite temperature (\( \beta J=50) \)
for the superlattice considered in Fig. 13 with \( h_{A}=h_{B}=h. \) Although
the central regions of each plateaux remain, as expected, the thermal fluctuations
suppress the quantum transitions.

\section{The Two-Spin Correlation Function in the Field Direction}

For sites in the subcell \( A \), for example, the two-spin static correlation
function in the field direction is defined by 
\begin{eqnarray}
\left\langle S_{j,m}^{A^{z}}S_{j+r,n}^{A^{z}}\right\rangle  & = & \left\langle a_{j,m}^{\dag }a_{j,n}a_{j+r,m}^{\dag }a_{j+r,n}\right\rangle +\nonumber \\
 &  & \nonumber \\
 & - & \frac{1}{2}\left( \left\langle a_{j,m}^{\dag }a_{j,m}\right\rangle +\left\langle a_{j,n}^{\dag }a_{j,n}\right\rangle \right) +\frac{1}{4}.
\end{eqnarray}
 The average value \( \left\langle a_{j,m}^{\dag }a_{j,n}a_{j+r,m}^{\dag }a_{j+r,n}\right\rangle  \)
can be obtained from the expression 
\begin{eqnarray}
\left\langle a_{j,m}^{\dag }a_{j,n}a_{j+r,m}^{\dag }a_{j+r,n}\right\rangle =\frac{1}{\pi }\int _{-\infty }^{\infty }\frac{Im\left( \left\langle \left\langle a_{j+r,n;}a_{j,m}^{\dag }a_{j,m}a_{j+r,n}^{\dag }\right\rangle \right\rangle \right) }{e^{\beta \omega }+1}d\omega , &  & \nonumber \\
 &  & 
\end{eqnarray}
 where the Fourier transform of the Green function \( \ll a_{j+r,n;}a_{j,m}^{\dag }a_{j,m}a_{j+r,n}^{\dag }\gg  \),
by using Wick`s theorem, can be shown to be written in the form 
\begin{eqnarray}
\left\langle \left\langle a_{j+r,n;}a_{j,m}^{\dag }a_{j,m}a_{j+r,n}^{\dag }\right\rangle \right\rangle  & = & \delta _{m,n}\delta _{r,0}\left\langle \left\langle a_{j+r,n;}a_{j,m}^{\dag }\right\rangle \right\rangle +\nonumber \\
 &  & \nonumber \\
+ &  & \left\langle a_{j,m}^{\dag }a_{j,m}\right\rangle \left\langle \left\langle a_{j+r,n;}a_{j+r,n}^{\dag }\right\rangle \right\rangle \nonumber \\
 &  & \nonumber \\
- &  & \left\langle a_{j+r,n}^{\dag }a_{j,m}\right\rangle \left\langle \left\langle a_{j+r,n;}a_{j,m}^{\dag }\right\rangle \right\rangle .
\end{eqnarray}
 Then, the correlation function \( \left\langle S_{j,m}^{A^{z}}S_{j+r,n}^{A^{z}}\right\rangle  \)
can be obtained from the equation 
\begin{eqnarray}
\left\langle S_{j,m}^{A^{z}}S_{j+r,n}^{A^{z}}\right\rangle  & = & \frac{\delta _{m,n}\delta _{r,0}}{\pi }\int _{-\infty }^{\infty }\frac{Im\left\langle \left\langle a_{j+r,n;}a_{j,m}^{\dag }\right\rangle \right\rangle }{e^{\beta \omega }+1}d\omega +\nonumber \\
 &  & \nonumber \\
- &  & \frac{\left\langle a_{j+r,n}^{\dag }a_{j,m}\right\rangle }{\pi }\int _{-\infty }^{\infty }\frac{Im\left\langle \left\langle a_{j+r,n;}a_{j,m}^{\dag }\right\rangle \right\rangle }{e^{\beta \omega }+1}d\omega +\nonumber \\
 &  & \nonumber \\
+ &  & \left\langle a_{j,m}^{\dag }a_{j,m}\right\rangle \left\langle a_{j+r,n;}a_{j+r,n}^{\dag }\right\rangle +\nonumber \\
 &  & \nonumber \\
- &  & \frac{1}{2}\left( \left\langle a_{j,m}^{\dag }a_{j,m}\right\rangle +\left\langle a_{j,n}^{\dag }a_{j,n}\right\rangle \right) +\frac{1}{4}.
\end{eqnarray}
 The Green function \( \ll a_{j+r,n};a_{j,m}^{\dag }\gg  \) can be obtained
by using eqs. (6), (13) and (18), and is given by {[}13{]}
\begin{eqnarray}
\left\langle \left\langle a_{j+r,n};a_{j,m}^{\dag }\right\rangle \right\rangle  & = & \frac{2}{NJ}\sqrt{\frac{n_{B}+1}{n_{A}+1}}\sum _{Q}\exp \left( -iqr\right) \left\{ f_{m,n}^{A}\left( \omega \right) \right. +\nonumber \\
 &  & \nonumber \\
 & + & \frac{1}{R_{Q}\left( \omega \right) }\left\{ \left( \left( f_{1,n_{B}}^{B}\left( \omega \right) \right) ^{2}-\left( f_{1,1}^{B}\left( \omega \right) \right) ^{2}\right) \right. \times \nonumber \\
 &  & \nonumber \\
 & \times  & \left[ f_{m,n_{A}}^{A}\left( \omega \right) f_{n,n_{A}}^{A}\left( \omega \right) f_{1,1}^{A}\left( \omega \right) -f_{m,n_{A}}^{A}\left( \omega \right) f_{1,n}^{A}\left( \omega \right) f_{1,n_{A}}^{A}\left( \omega \right) \right. +\nonumber \\
 &  & \nonumber \\
 & + & \left. f_{1,m}^{A}\left( \omega \right) f_{1,n}^{A}\left( \omega \right) f_{1,1}^{A}\left( \omega \right) -f_{1,m}^{A}\left( \omega \right) f_{n,n_{A}}^{A}\left( \omega \right) f_{1,n_{A}}^{A}\left( \omega \right) \right] +\nonumber \\
 &  & \nonumber \\
 & + & f_{1,n_{B}}^{B}\left( \omega \right) \left[ f_{1,n}^{A}\left( \omega \right) f_{m,n_{A}}^{A}\left( \omega \right) \exp \left( iq\right) +f_{n,n_{A}}^{A}\left( \omega \right) f_{1,m}^{A}\left( \omega \right) \exp \left( -iq\right) \right] +\nonumber \\
 &  & \nonumber \\
 & + & \left. \left. f_{1,1}^{B}\left( \omega \right) \left[ f_{m,n_{A}}^{A}\left( \omega \right) f_{n,n_{A}}^{A}\left( \omega \right) +f_{1,m}^{A}\left( \omega \right) f_{1,n}^{A}\left( \omega \right) \right] \right\} \right\} .
\end{eqnarray}
 Therefore, introducing the previous result in eq. (56) and by using eq. (47)
we can obtain numerically the direct static correlation function \( \left\langle S_{j,m}^{A^{z}}S_{j+r,n}^{A^{z}}\right\rangle - \)\( \left\langle S_{j,m}^{A^{z}}\right\rangle \left\langle S_{j+r,n}^{A^{z}}\right\rangle  \),
and the results are shown in Figs. 15 and 16.

In Fig. 15 we present the direct static correlation function as a function of
\( r, \) at \( T=0, \) for two values of the field of the transverse field,
\( h=0.693 \) and \( h=0.75, \) and for \( J=1 \), \( J_{A}=2 \), \( J_{B}=3, \)
\( h_{A}=h_{B}=h \) and \( n_{A}=n_{B}=2, \) which are the same set of parameters
of Fig. 12. The values of the critical fields, for this set of parameters, are
given in eq. (52) and the first two values are \( h_{1c}\cong 0.691 \) and \( h_{2c}\cong 1.096. \)
As can be verified in Fig. 15, for these values of the field\( , \) the correlation
function presents an oscillatory behaviour and its period increases ( period
\( \rightarrow \infty ) \) as the field aproaches a critical value (\( h\rightarrow h_{c}). \)

In Fig. 16 it is also presented the direct static correlation function for the
same set of lattice parameters of Fig. 15, for \( h=1.091 \) and for \( h=1.1. \)
Although \( h=1.091, \) which lies in the region of increasing magnetization
(see Fig. 11), and is very close to the critical field \( h_{2c}\cong 1.096, \)
the oscillatory behaviour is still present in the correlation function. On the
other hand, when \( h=1.1 \), which lies in a plateaux of the magnetization
(see also Fig. 11), there is no oscillatory behaviour. It means that the period
of the oscillation tends to infinity, and this result is still valid for any
value of the field in the plateaux. This is consistent with the scaling form
and the analytical continuation proposed for this correlation function {[}15{]},
where the correlation length is associated to the oscillation period.

Although the direct correlations \( \left\langle S_{j,m}^{B^{z}}S_{j+r,n}^{B^{z}}\right\rangle  \)
\( - \) \( \left\langle S_{j,m}^{B^{z}}\right\rangle \left\langle S_{j+r,n}^{B^{z}}\right\rangle  \)
and \( \left\langle S_{j,m}^{A^{z}}S_{j+r,n}^{B^{z}}\right\rangle  \) \( - \)
\( \left\langle S_{j,m}^{A^{z}}\right\rangle \left\langle S_{j+r,n}^{B^{z}}\right\rangle  \)
are not presented, they are easily obtained from the shown results and exhibit
qualitatively the same behaviour as \( \left\langle S_{j,m}^{A^{z}}S_{j+r,n}^{A^{z}}\right\rangle - \)\( \left\langle S_{j,m}^{A^{z}}\right\rangle \left\langle S_{j+r,n}^{A^{z}}\right\rangle  \).
This is a consequence of the fact that the critical fields are not dependent
on the subcell.

\section{Conclusions}

We have considered the XY (\( s=1/2 \)) model on the alternating one-dimen\-sional
superlattice (closed chain), and an exact solution was obtained by using the
Green function method. The excitation spectrum was determined, and explicit
expressions were obtained at arbitrary temperature for the internal energy,
the specific heat, the magnetization, the susceptibility, and the two-spin static
correlation function in the field direction.

The specific heat as a function of temperature, depending on the superlattice
parameters, can present a single or a double peak, and we have shown that, at
\( T=0 \), \( dC/dT \) is different from zero provided there is a zero energy
mode on the spectrum.

In the \( T=0 \) limit, the behaviour of the system was studied as a function
of the transverse field , and we have shown that the induced magnetization as
a function of the field presents, alternately, regions of plateaus and of variable
magnetization. Also in this temperature limit, the susceptibility in the field
direction, \( \chi ^{zz} \), presents singularities which are associated to
phase transitions of second kind induced by the field. These critical points
are consequence of the presence of zero energy modes with wave number \( 0 \)
or \( \pi  \). In passing, it should be noted that this critical behaviuor,
as expected, is suppresed at finite temperatures.

These transitions have been treated within the real space renormalization group
approach {[}16{]}, and its critical exponents determined. The critical exponents
can be also obtained directly from the exact expression, as shown elsewhere
{[}13{]}.

The two-spin static correlation function in the field direction, as a function
of the separation between the spins, presents an oscillating behaviour in the
regions where the magnetization is not constant, and the period of oscillations
increases as the field approaches the critical value, and diverges at \( h=h_{c}. \)
This behaviour confirms that the static correlation function satisfies the analytical
extension of the scaling form proposed for the homogeneous case {[}15{]}.

Finally we would like to point out that the magnetization as a function of the
field has qualitatively the same behavior as those experimentally obtained for
the NdCu\( _{2} \) in the low temperature limit, and these results have been
obtained recently by Ellerby \textit{et al}. {[}17{]} and Loewenhaupt \textit{et
al.} {[}18{]}. The agreement is more remarkable in the very low temperature
limit, since in this limit the structure is analogous to a superlattice. Although
this material is described by a Heisenberg type Hamiltonian, this result suggests
that the lattice structure is a predominant factor in defining the magnetic
properties of the material.

\section{References}

\noindent {[}1{]} J. R. Childress, R. Kergoat, O. Durand, J.M. George, P. Galtier,
J. Miltat and A. Shulhl, \textit{J. Magn. \& Magn. Mater.} \textbf{130} (1994)
13.

\noindent {[}2{]} B. V. Reddy, S. N. Khanna, P. Jena, M. R. Press and S. S.
Jaswal, \textit{J. Magn. \&} \textit{Magn. Mater.} \textbf{130} (1994) 255.

\noindent {[}3{]} J. J. Rhyne, M. B. Salomon, C. P. Flynn, R. W. Erwing and
J. A. Borchers, \textit{J. Magn. \&} \textit{Magn. Mater.} \textbf{129} (1994)
39.

\noindent {[}4{]} E. Lieb, T. Schultz and D. C. Mattis, \textit{Ann. Phys.}(NY)
\textbf{16} (1961) 406.

\noindent {[}5{]} J. Stolze, A. Nopert and G. M\"{u}ller, \textit{Phys. Rev.}
\textbf{B52} (1995) 4319.

\noindent {[}6{]} V. M. Kontorovich and V. M. Tsurkenik, \textit{Sov. Phys.
JETP} \textbf{26} (1968) 687.

\noindent {[}7{]} J. H. H. Perk, H. W. Capel and M. J. Zuilhof , \textit{Physica}
\textbf{81A} (1975) 319.

\noindent {[}8{]} J. H. H. Perk, Th. J. Siskens and H. W. Capel, \textit{Physica}
\textbf{89A} (1977) 304.

\noindent {[}9{]} L. L. Gon\c{c}alves and J.P. de Lima, \textit{J. Magn. Magn.
Mater.} \textbf{140-144} (1995) 1606.

\noindent {[}10{]} J. P. de Lima and L.L. Gon\c{c}alves, \textit{Mod. Phys.
Lett.} \textbf{B22} (1996) 1077.

\noindent {[}11{]} L. L. Gon\c{c}alves, \textit{Theory of Properties of Some
One-Dimensional Systems,} (D.Phil. Thesis, University of Oxford, 1977).

\noindent {[}12{]} D. N. Zubarev, \textit{Sov. Phys. Usp.} \textbf{3} (1960)
320.

\noindent {[}13{]} J. P. de Lima, XY-Model in One-Dimension: New Exact Results,
(PhD Thesis, Universidade Federal do Cear\'{a}, 1997) in portuguese.

\noindent {[}14{]} E. R. Smith, \textit{J. Phys.} \textbf{C3} (1970) 1419.

\noindent {[}15{]} J. P. de Lima and L. L. Gon\c{c}alves, \textit{Mod. Phys.
Lett.} \textbf{B14\&15} (1994) 871.

\noindent {[}16{]} L. L. Gon\c{c}alves and J.P. de Lima, \textit{J. Phys.:
Condens. Matter} \textbf{9} (1997) 3447.

\noindent {[}17{]} M. Ellerby, K. A. McEwen, M. de Podesta, M. Rotter and E.
Gratz E., \textit{J. Phys.: Condens. Matter} \textbf{7} (1995) 1897.

\noindent {[}18{]} M. Loewenhaupt, Th. Reif, P. Svoboda, S. Wagner, M. Wafenschimidt,
H. V. Lohneysen, E. Gratz, M.Rotter, B. Lebech and Th. Haub, \textit{Z. Phys.}
\textbf{B101} (1996) 499.

\section{Acknowledgements}

The partial financial support from CNPq and Finep (Brazilian Agencies) is gratefully
acknowledged.

\section{Figure Captions}

\vspace{0in}

\noindent \textbf{Fig. 1}- Unit cell of the alternating superlattice.

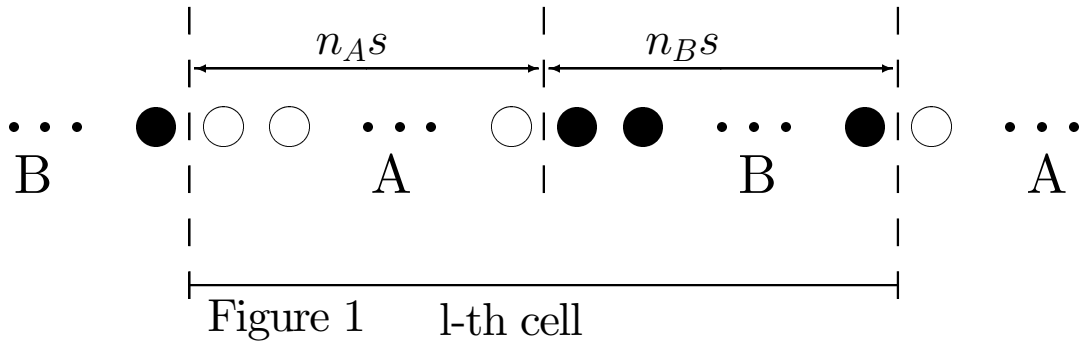
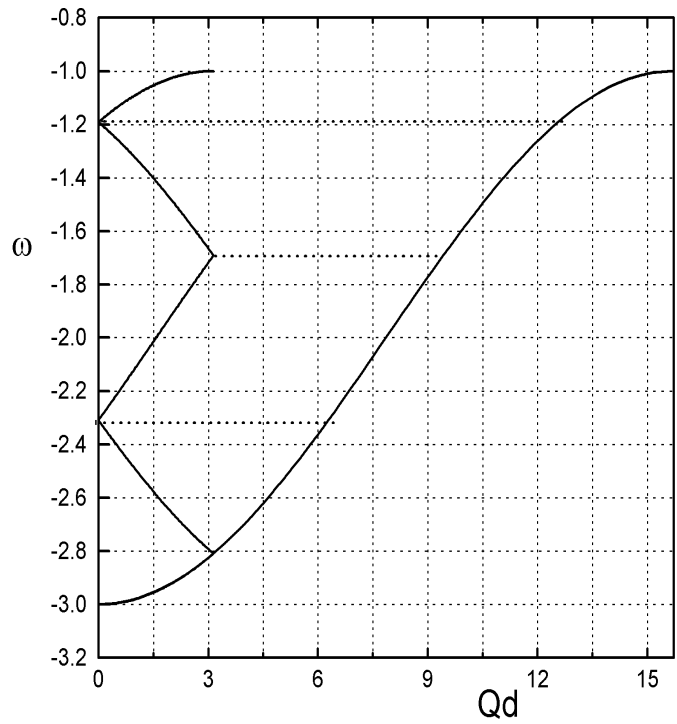
\noindent \textbf{Fig. 2}- Excitation spectrum of the homogeneous lattice for
\( n_{A}=2,n_{B}=3, \) \( J= \) \( J_{A}=, \) \( J_{B}=1, \) and \( h_{A}=h_{B}=h=2. \)

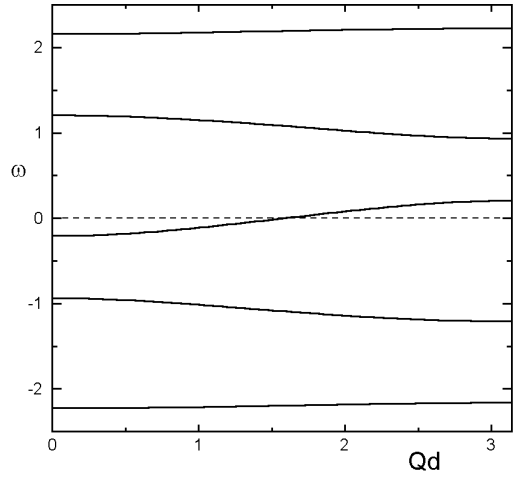
\noindent \textbf{Fig. 3}- Excitation spectrum for \( n_{A}=2,n_{B}=3, \) \( J=1, \)
\( J_{A}=2, \) \( J_{B}=3, \) and \( h_{A}=h_{B}=0. \)

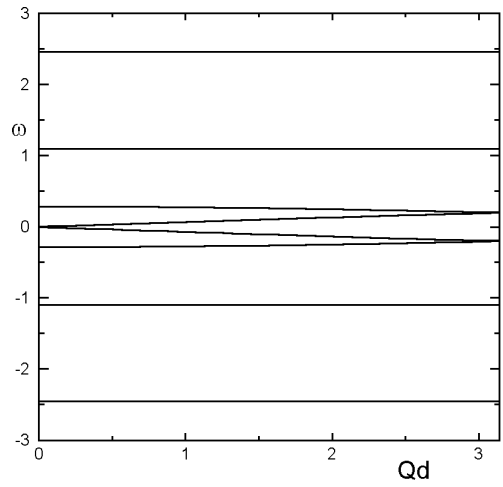
\noindent \textbf{Fig. 4}- Excitation spectrum for \( n_{A}=n_{B}=4, \) \( J=1, \)
\( J_{A}=0.3, \) \( J_{B}=3, \) and \( h_{A}=h_{B}=0. \)

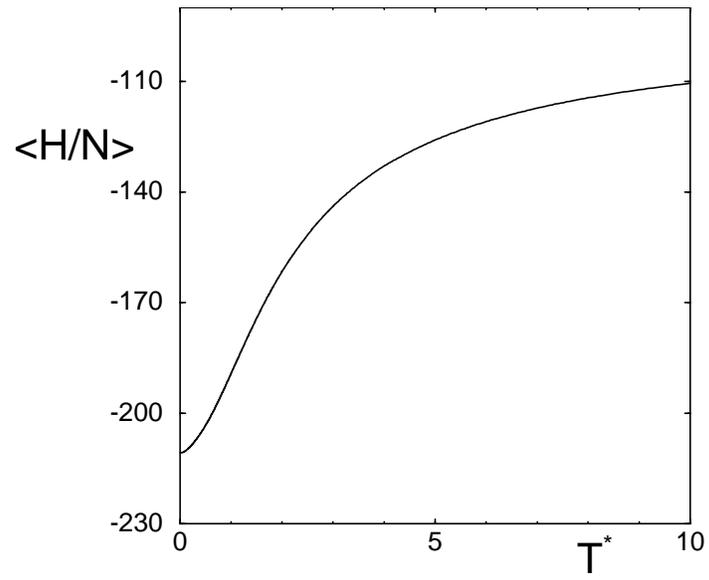
\noindent \textbf{Fig. 5}- Internal energy as a function of temperature \( (T^{*}\equiv k_{B}T) \)
for \( n_{A}=n_{B}=10, \) \( J=1, \) \( J_{A}=2, \) \( J_{B}=3, \) and \( h_{A}=h_{B}=1.5. \)

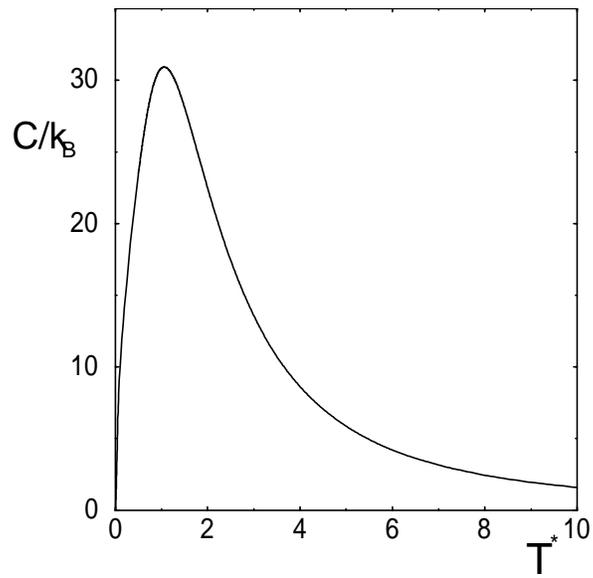
\noindent \textbf{Fig. 6}- Specific heat as a function of temperature \( (T^{*}\equiv k_{B}T) \)
for \( n_{A}=n_{B}=10, \) \( J=1, \) \( J_{A}=2, \) \( J_{B}=3, \) and \( h_{A}=h_{B}=1.5. \)

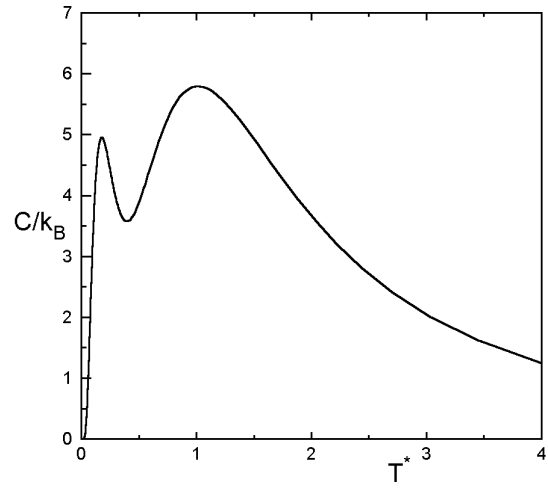
\noindent \textbf{Fig. 7}- Specific heat as a function of temperature \( (T^{*}\equiv k_{B}T) \)
for \( n_{A}=n_{B}=2, \) \( J=1, \) \( J_{A}=2, \) \( J_{B}=3, \) and \( h_{A}=h_{B}=1.3. \)

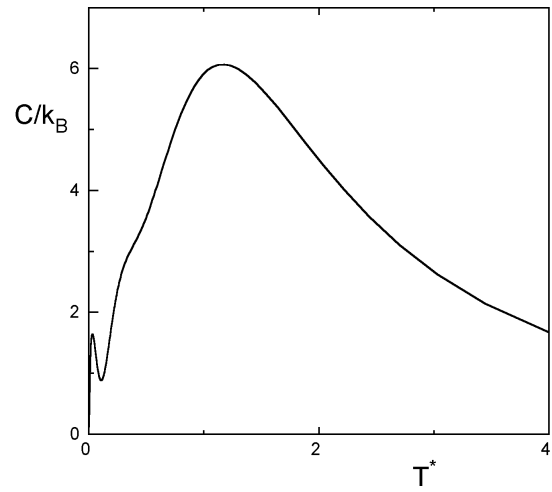
\noindent \textbf{Fig. 8}- Specific heat as a function of temperature \( (T^{*}\equiv k_{B}T) \)
for \( n_{A}=n_{B}=2, \) \( J=1, \) \( J_{A}=2, \) \( J_{B}=3, \) and \( h_{A}=h_{B}=1.75. \)

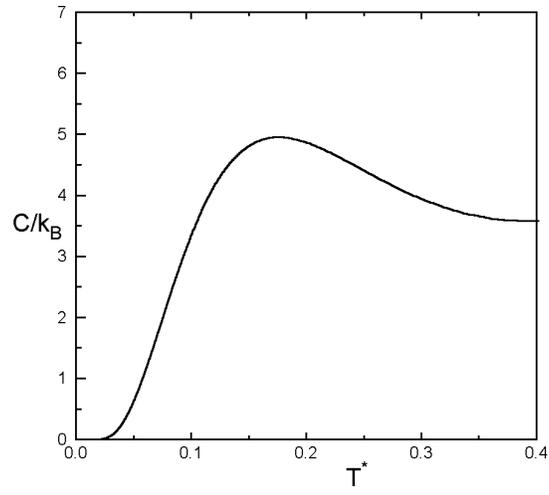
\noindent \textbf{Fig. 9}-The low temperature behaviour of the specific heat
shown in Fig. 7 \( (T^{*}\equiv k_{B}T). \)

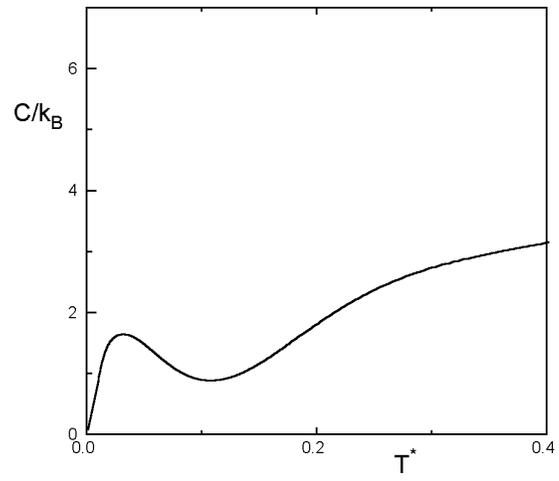
\noindent \textbf{Fig. 10}- The low temperature behaviour of the specific heat
shown in Fig. 8 \( (T^{*}\equiv k_{B}T). \)

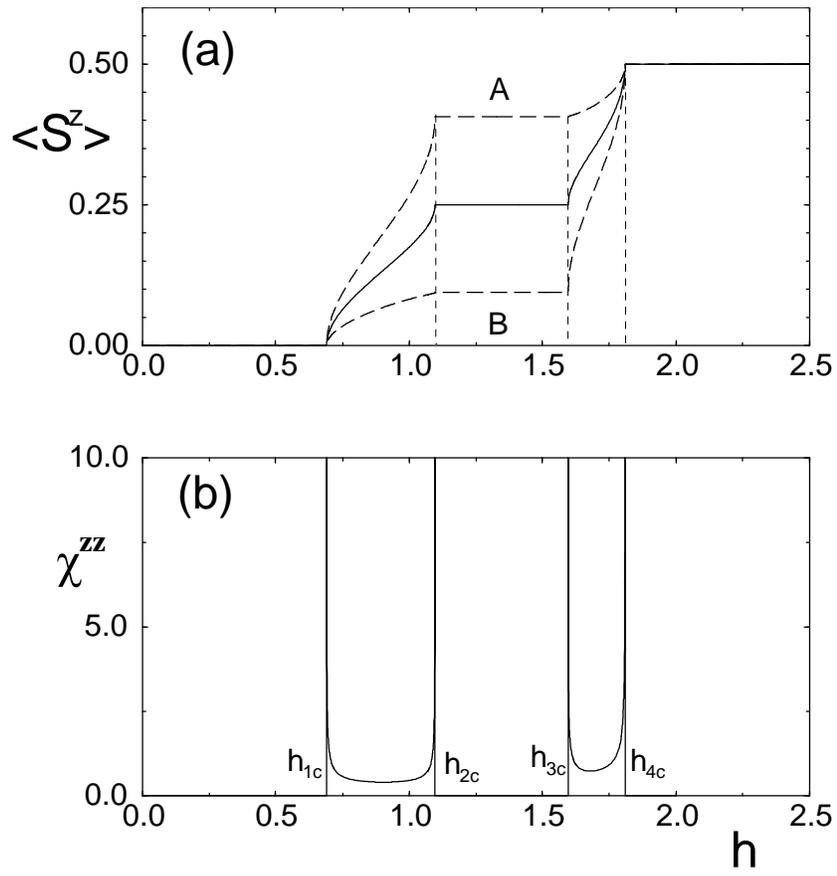
\noindent \textbf{Fig. 11}- (a) Average magnetization per subcell A and B (dashed
line), and per unit cell (continuous line) as a function of the field, and (b)
the susceptibility in the field direction also as a function of the field, for
\( n_{A}=n_{B}=2, \) \( J=1, \) \( J_{A}=2, \) \( J_{B}=3, \) and \( h_{A}=h_{B}=h, \)
at \( T=0 \). The critical fields are \( h_{1c}\cong 0.691, \) \( h_{2c}\cong 1.096, \)
\( h_{3c}\cong 1.5960 \) and \( h_{4c}\cong 1.809. \)

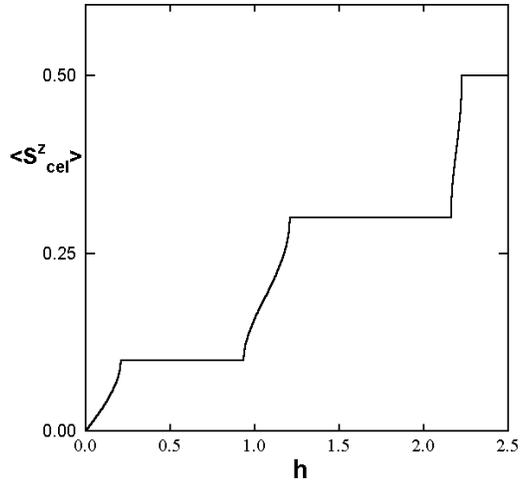
\noindent \textbf{Fig. 12}- Average magnetization per unit cell as a function
of the field, for \( n_{A}=2,n_{B}=3, \) \( J=1, \) \( J_{A}=2,J_{B}=3, \)
and \( h_{A}=h_{B}=h, \) at \( T=0. \)

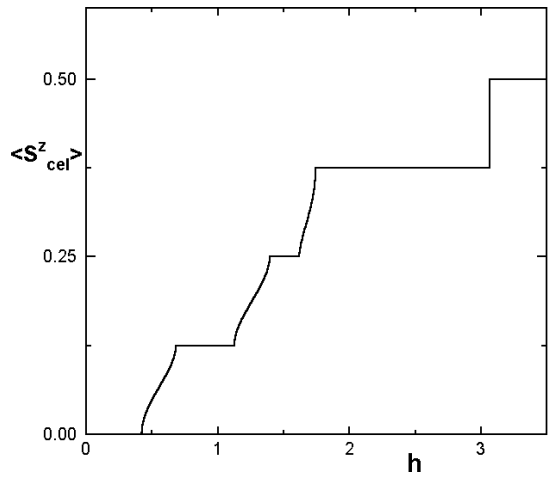
\noindent \textbf{Fig. 13}- Average magnetization per unit cell as a function
of the field, for \( n_{A}=n_{B}=4, \) \( J=1, \) \( J_{A}=2,J_{B}=3, \)
and \( h_{A}=1.25h_{B}=h, \) at \( T=0. \)

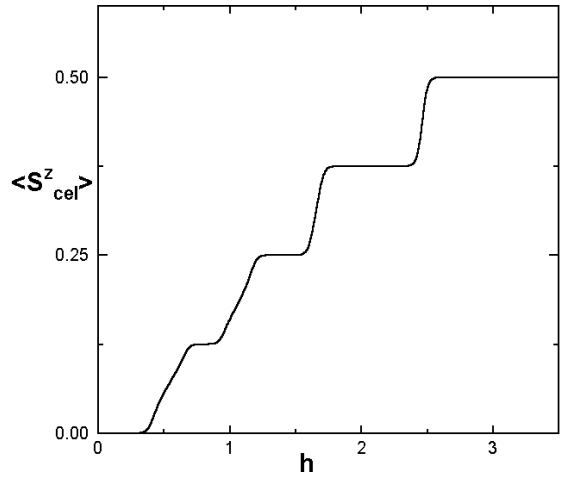
\noindent \textbf{Fig. 14}- Average magnetization per unit cell as a function
of the field, for \( n_{A}=n_{B}=4, \) \( J=1, \) \( J_{A}=2,J_{B}=3, \)
and \( h_{A}=h_{B}=h, \) at finite temperature (\( \beta J=50) \).

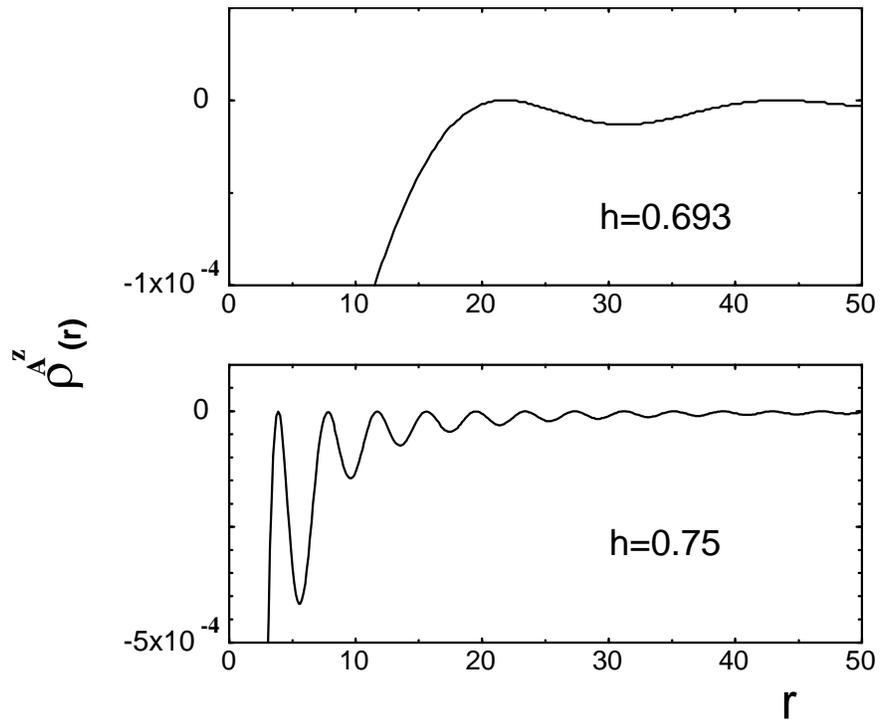
\noindent \textbf{Fig. 15}- The direct static correlation function in the field
direction, \( \rho ^{A^{z}}(r)\equiv \left\langle S_{j,2}^{A^{z}}S_{j+r,2}^{A^{z}}\right\rangle -\left\langle S_{j,2}^{A^{z}}\right\rangle ^{2}, \)
as a function of the distance between cells, for \( n_{A}=n_{B}=2, \) \( J=1, \)
\( J_{A}=2, \) \( J_{B}=3, \) and \( h_{A}=h_{B}=h, \) at \( T=0 \), for
\( h=0.75 \)(a) and \( h=0.693 \)(b).

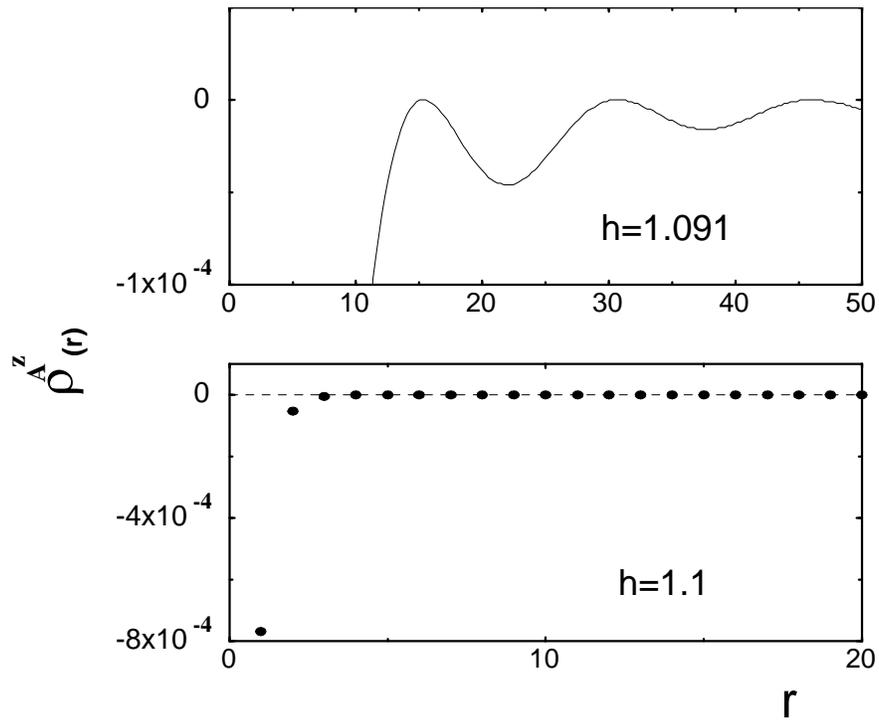
\noindent \textbf{Fig. 16}- The direct static correlation function in the field
direction, \( \rho ^{A^{z}}(r)\equiv \left\langle S_{j,2}^{A^{z}}S_{j+r,2}^{A^{z}}\right\rangle -\left\langle S_{j,2}^{A^{z}}\right\rangle ^{2}, \)
as a function of the distance between cells, for \( n_{A}=n_{B}=2, \) \( J=1, \)
\( J_{A}=2, \) \( J_{B}=3, \) and \( h_{A}=h_{B}=h, \) at \( T=0 \), for
\( h=1.1 \)(a) and \( h=1.091 \)(b).

\end{document}